\documentstyle[twocolumn,graphics,prl,aps]{revtex}

\newcommand{\lu}{LuNi$_2$B$_2$C}

\begin{document}

\wideabs{
\draft
\title{Delocalized quasiparticles throughout the vortex state in 
       $s$-wave superconductor LuNi$_2$B$_2$C}

\author{Etienne Boaknin, R.W. Hill,
Cyril Proust, C. Lupien and Louis Taillefer}

\address{Canadian Institute for Advanced Research}

\address{
Department of Physics, University of Toronto, Toronto,
Ontario M5S 1A7, Canada}

\author{P.C. Canfield}

\address{Ames Laboratory, Department of Physics and Astronomy, Iowa
  State University, Ames, Iowa 50011}

\date{\today}

\maketitle

\begin{abstract}
  Quasiparticle transport in the vortex state of an $s$-wave
  superconductor at $T \to 0$ was investigated by measuring the
  thermal conductivity of LuNi$_2$B$_2$C down to 70 mK in a magnetic
  field perpendicular to the heat current.  In zero field, there is no
  electronic conduction, as expected for a superconducting gap without
  nodes. However, as soon as vortices enter the sample quasiparticles
  are seen to conduct remarkably well, even better than they would in
  a typical $d$-wave superconductor. This is in stark conflict with
  the widely held view that quasiparticle states in $s$-wave
  superconductors just above $H_{c1}$ should be localized and bound to
  the vortex core.

\end{abstract}

\pacs{PACS numbers: 74.70.Tx, 74.25.Fy }}

The physics of weakly interacting electrons in non-magnetic weakly
disordered metals and superconductors with three-dimensional crystal
structures is thought to be extremely well understood, in the
following limits: the $s$-wave superconductor in zero magnetic field
and the metallic state at low temperature.  The former is described by
the theory of Bardeen, Cooper and Schrieffer (BCS) and the latter by
the Fermi-liquid theory of Landau, two cornerstones of solid state
theory.  These two limits can be achieved in the same material and the
electron system can be made to go smoothly from one to the other
simply by applying a magnetic field $H$, going from zero to above
$H_{c2}$, the upper critical field beyond which superconductivity is
destroyed.

The $s$-wave superconductor at $T \to 0$ and $H=0$ is a thermal
insulator, in that the electron system cannot transport any heat: the
Cooper pairs carry no entropy and there are no quasiparticles. On the
other hand, the normal state above $H_{c2}$ is a thermal metal. The
two states may be characterized by their heat conductivity
$\kappa(T)$, through the value of the residual linear term 
$\kappa_0/T$, defined as
the limit of $\kappa/T$ when $T \to 0$.  $\kappa_0/T$ is zero in the
$s$-wave superconductor and finite in the metal, where it is given
precisely by the Wiedemann-Franz law. This law is a universal relation
between charge conductivity and heat conductivity at $T \to 0$, given
by $\kappa_0/T = L_0 / \rho_0$, where $\rho_0$ is the residual
electrical resistivity.

A basic question arises: how does the electron system evolve between
these two well-defined and well-understood limits, these two distinct
states of matter?  Remarkably, this question is still largely
unanswered.  

In superconductors of the second kind (type II),
the electron system
responds to an applied magnetic field greater than a critical value
$H_{c1}$ by generating vortices, {\it i.e.} lines of quantized flux at
the core of which the superconducting gap goes to zero.  Our basic
question then becomes: what is the nature of the electronic states as
a function of vortex density? The density is given strictly by the
magnetic field strength $B$ as $\Phi_0/B$, where $\Phi_0=hc/2e$ is the
quantum of flux.  Considerable progress has been made over the past
decade in elucidating the extreme limits of very low density and very
high density. A single isolated vortex in an $s$-wave superconductor
with an isotropic zero-field gap $\Delta_0$ introduces a discrete set of energy
levels inside the vortex core, the lowest level lying an energy
$\Delta_0/2 k_F \xi$ above the Fermi energy, with higher levels
spaced equally by twice that amount \cite{Caroli}, where $k_F$ is the
Fermi wavevector and $\xi$ is the vortex core radius. The density of
states associated with these levels is roughly that of the normal
state, but limited to an area $\pi \xi^2$, where $\xi$ is of order
100-1000 \AA ~in most $s$-wave superconductors. 
The important point is that the states associated with a single vortex
are not extended but localized, bound to the vortex core.  The
detailed variation of the local density of states associated with a
vortex was mapped out using STM on NbSe$_2$ by Hess {\it et al.}
\cite{Hess} and interpreted in terms of states bound to
the vortex core \cite{theory}.  In the presence of a vortex lattice,
the quasiparticle states form bands \cite{Norman,Yasui}, but at fields
low enough that the intervortex separation $d$ is much larger than the
vortex core size, the low-lying bands are very flat \cite{Yasui},
essentially identical to the localized levels of the isolated vortex
(and therefore also gapped).
In other words, when the ratio $\xi/d$ is very small, the tunneling
between states on neighbouring vortices is exponentially small,
causing very little dispersion.

In the other limit of fields close to $H_{c2}$, 
quasiparticles are delocalized, as inferred for example from the
observation of de Haas-van Alphen oscillations in type-II
superconductors such as NbSe$_2$ \cite{dHvA-NbSe2}, V$_3$Si
\cite{dHvA-V3Si} and YNi$_2$B$_2$C \cite{dHvA-boro}.  In
this limit, calculated bands are broadened Landau levels with
zero-energy quasiparticles occuring at several $k$ points in the
magnetic Brillouin zone \cite{Norman}.  Te\u{s}anovi\'{c} and
Sacramento have argued that the high-field gapless phase is separated from
the low-field gapped phase by a quantum level crossing transition
\cite{Tesanovic}.

The picture for very low and very high vortex densities is therefore
expected to be much the same as at $H=0$ and $H=H_{c2}$, namely
that of localized vs delocalized quasiparticles,
respectively.  The best way to investigate the nature of quasiparticle
states and the possibility of a metal-insulator transition is to
measure heat transport. Note that charge transport is much less ideal, 
either theoretically since charge is not a conserved quantity, or experimentally since
charge dynamics is dominated by the dissipation caused by moving vortices
(flux flow regime).
In this Letter, we report on the
first systematic study of heat transport throughout the vortex state of
an $s$-wave superconductor. 
Contrary to all expectations, we observe highly delocalized quasiparticles at
temperatures as low as $T_c/200$ for all values of the field in the vortex state.  
There is no indication of an insulator-to-metal 
transition between $H_{c1}$ and $H_{c2}$, eventhough the ratio
$d/\xi$ starts off as large as 10-20 at $H_{c1}$.

The borocarbide material LuNi$_2$B$_2$C was chosen because of its
large critical field ratio $H_{c2}/H_{c1}$ $\simeq$~100, which
allows access to a wide range of intervortex separations.  It is a
non-magnetic superconductor 
thought to be
characterized by $s$-wave pairing and a gap without nodes.
Its crystal structure is tetragonal and its Fermi surface properties
nearly isotropic.  The thermal conductivity
was measured in a dilution refrigerator using a standard steady-state
technique.  The temperature was swept at fixed magnetic fields, from 70
mK up, in fields ranging from 0 to 8 T.  The field was applied
perpendicular to the heat current and parallel to the $c$-axis ([001]
direction).  Below 1.5 T, the sample was
cooled in the field to ensure a good field homogeneity and a
well-ordered vortex lattice.  
Above 1.5 T, the results were independent of cooling procedure.  The
single crystal was grown by a melting flux method described
elsewhere \cite{Canfield}.  The sample was shaped and polished into
a rectangular parallelepiped of width 0.495 mm (along [010]) and
thickness 0.233 mm (along [001]), with a 1.59 mm separation between
contacts (along [100]).
The same contacts were used for electrical resistivity and thermal
conductivity measurements, made with non-superconducting solder and
yielding contact resistances lower than 5 m$\Omega$ at 300 K.  In zero
magnetic field, the resistivity was $\rho$(300 K) = 35 $\mu \Omega$ cm
at room temperature and $\rho_0$ = 1.30~$\mu \Omega$ cm at $T \to 0$,
using an extrapolation of the form $\rho$ = $\rho_0$ + AT$^2$ between
$T_c$ and 50 K, yielding an RRR of 27.  $T_c$ was found to be 16 K
with a transition width of 1~K.  Magnetoresistive measurements
performed at 1.8 K yielded 
$H_{c2}$ = 7 T and
$\rho$ = 1.67 $\mu \Omega$ cm just above $H_{c2}$.
The zero-temperature coherence length $\xi_0 = 70$~ \AA, using
$H_{c2}(0) = \Phi_0 / 2 \pi \xi_0^2$. 
\begin{figure}
\resizebox{\linewidth}{!}{\includegraphics{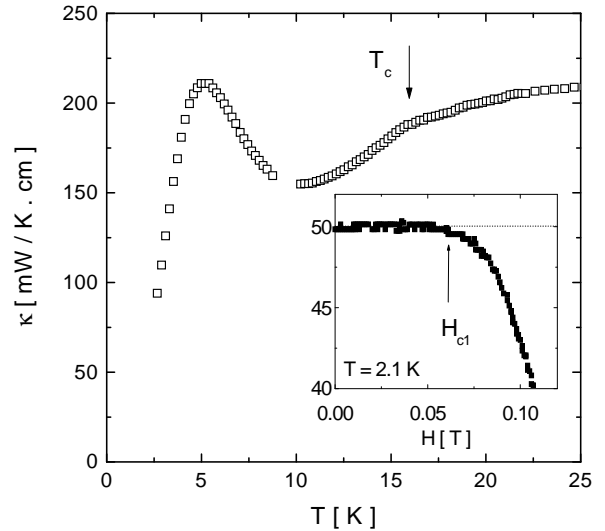}} 
\caption{Thermal conductivity of \lu\ between 2.5 and 25~K. The  
  superconducting transition occurs at $T_c$ = 16~K and can be seen as
  a kink in the thermal conductivity. The large peak centered at 5~K
  is due to a rapid increase in the phonon mean free path as the
  electrons condense below $T_c$.  {\it Inset}: Thermal conductivity
  at $T$ = 2.1~K as a function of magnetic field. The downward
  deviation from a constant conductivity at a field $H_{c1}
  \simeq 60$~mT
  signals the onset of vortex entry into the sample.}
\end{figure}
Note that the radius of the vortex cores
are somewhat larger than this at low fields (by a factor 2 or so) as
measured by muon spin rotation ($\mu$SR) in this compound
\cite{Price}. A similar expansion was previously reported in NbSe$_2$
\cite{Sonier}.

The thermal conductivity of \lu\ in zero magnetic field is plotted in
Fig.~1 as $\kappa(T)$ vs $T$ up to $T_c$ and beyond. The total
conductivity is the sum of an electronic and a phononic contribution:
$\kappa = \kappa_e + \kappa_{ph}$.  Above $T_c$, $\kappa$ is dominated
by electrons, as $\kappa_{ph}$ is severely limited by electron
scattering. As the temperature is decreased below $T_c$, two things
happen as electrons condense into Cooper pairs: $\kappa_e$ decreases
as there are fewer and fewer thermally excited quasiparticles and
$\kappa_{ph}$ increases as fewer unpaired electrons are available to
scatter phonons.  The first effect is seen as a dip right at $T_c$ and
the latter leads to a huge increase in phonon mean free path giving
rise to the pronounced peak at 5~K. This increase is expected to stop
when the number of thermally excited quasiparticles becomes
vanishingly small. In a typical BCS superconductor such as Nb, the
peak occurs roughly at $T_c$/5, the temperature at which $\kappa_e$
becomes negligible (less than 1-2\% of its normal state value).  The
fact that in \lu\ the peak is even higher in temperature is an
indication that there are very few quasiparticles below $T_c$/3, which
tends to confirm the $s$-wave nature of the gap and rule out any
sizable gap anisotropy.  At temperatures below $T_c$/3, $\kappa$ is
essentially all due to phonons, whose mean free path eventually grows
to reach the size of the crystal, at which point a cubic temperature
dependence is observed. This is shown in Fig.~2, in a plot of
$\kappa/T$ vs $T^2$, which allows us to easily separate the electronic
term linear in $T$ from the phononic term cubic in $T$.  The
zero-field curve is indeed linear in such a plot (below 150 mK), with
zero intercept and a slope in quantitative agreement with the known
sound velocities and sample dimensions, as reported earlier
\cite{Boaknin}.  The absence of any residual linear term is a second
confirmation that there are no nodes in the gap (at least in the basal
plane).  To be more specific, the data sets an upper limit of
$\kappa_0/T \simeq 0.01$~mW~K$^{-2}$~cm$^{-1}$ $\simeq
(\kappa_N/T)(1/2000)$ on a possible residual linear term arising from
nodes in the basal plane. Theoretically, the value expected from a
line node in the basal plane is roughly $\kappa_0/T \simeq
(\kappa_N/T)(\hbar \Gamma / 2 \Delta_0)$, where $\Gamma = 1/2\tau_N$
is the impurity scattering rate and $\Delta_0$ is the gap maximum
\cite{Graf}.  We can estimate $\tau_N$, the transport relaxation time
in the normal state, from kinetic theory: $\kappa_N/T = L_0/\rho_0 =
\frac{1}{3} \gamma_N v_F^2 \tau_N$, where $\gamma_N$ is the Sommerfeld
coefficient of the specific heat and $v_F$ is the Fermi velocity.  The
measured values are $\gamma_N = 19.5$~mJ~K$^{-2}$~mole$^{-1}$
\cite{Michor}, $\Delta_0 = 3.2~k_B T_c$ \cite{Delta} and $v_F \simeq 2
\times 10^5$~m/s \cite{Kogan}. These numbers give a mean free path
$l_N = v_F \tau \simeq 500$~\AA, and $\hbar \Gamma \simeq 0.5
\Delta_0$, from which we would expect $\kappa_0/T \simeq \frac{1}{4}
\kappa_N/T$ if there were nodes in the gap. This huge discrepancy, by
a factor of nearly three orders of magnitude, argues strongly against
nodes in the gap function of \lu\ (and the insensitivity to impurities
argues against an unconventional order parameter).

As seen in Fig.~2, a residual linear term is induced by a magnetic field
applied perpendicular to the heat current.
When the field exceeds
$H_{c2}(0) = 7$~T, the conductivity is dominated by electrons and
$\kappa/T \simeq \kappa_e/T$ is essentially constant, as seen for 8 T,
with a value in perfect agreement with the Wiedemann-Franz
law: $\kappa_e/T =L_0/\rho$, where $\rho$ = $\rho$(8~T).
Turning to very low fields, we use the fact that
phonons are scattered by vortices to determine {\it in
situ} the field $H_{c1}$ at which vortices first enter the sample.
This is shown in the inset of Fig.~1, where $\kappa_{ph}$ at $T=2.1$~K
is seen to suddenly deviate from its field-independent value at $H_{c1}
\simeq 60$~mT.  
Note that at 60 mT, it is known from small-angle neutron scattering \cite{neutrons}
that vortices form a well-ordered square lattice in LuNi$_2$B$_2$C.
The question of fundamental interest is: 
how is the electronic conduction
modified by the presence of a vortex lattice?  
The low field behavior is
shown in the lower panel of Fig.~2. The curves of $\kappa/T$ vs $T^2$
for fixed fields of 50, 75, 100, 200 and 300 mT are seen to be rigidly
shifted relative to the zero field curve.
The growth of the corresponding residual linear term, which is the
electronic contribution, is plotted in Fig.~3. The striking 
\begin{figure}
\resizebox{\linewidth}{!}{\includegraphics{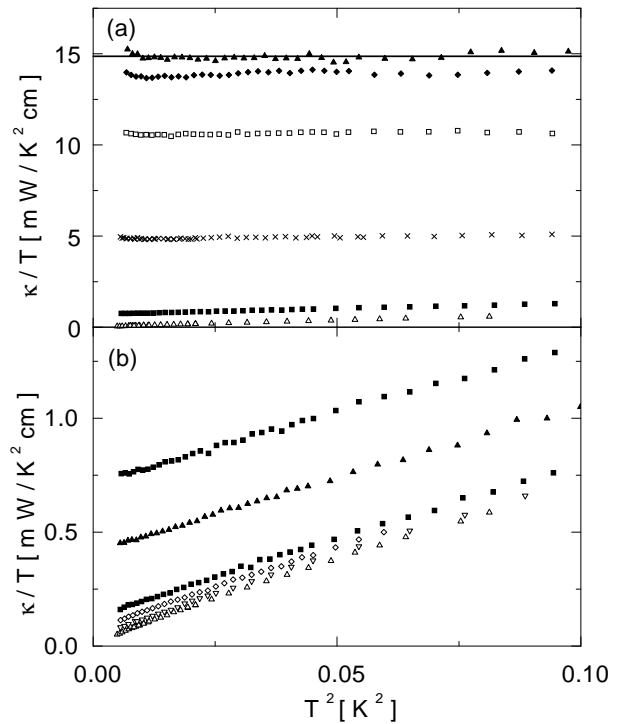}}
\caption{Temperature dependence of the thermal conductivity at several
  applied fields in the range 70~mK $< T < 300$~mK. The data is
  plotted as $\kappa/T$ vs $T^2$ to separate the electronic and
   phononic contributions. Fig~1 (a) shows the data for H = 0,
  0.3, 1.5, 4, 6 and 8 T, in increasing order. The solid line indicates the
  value expected from the Wiedemann-Franz law above $H_{c2}$. Fig.~1
  (b) shows the data for H = 0, 50, 75, 100, 200 and 300~mT, in
  increasing order.  At low fields, 
  the phononic term remains largely unchanged as the electronic
  term grows with increasing field.  }
\end{figure}
feature is
that this growth starts right at $H_{c1}$, as can be 
seen from the inset. Whereas quasiparticles in an $s$-wave superconductor
close to $H_{c1}$ are
theoretically expected to be localized and thus incapable of
transporting heat at $T \to 0$, the observed quasiparticle
conduction is excellent.  
\lu\ is therefore a thermal metal (with fully delocalized quasiparticles)
throughout the vortex state, at least down to $T_c/200$. 
This provides a natural explanation for the fact that de Haas-van Alphen
oscillations in YNi$_2$B$_2$C persist down to unusually low fields 
($H_{c2}/5$) \cite{dHvA-boro} and no localized core states have been detected in STM 
spectroscopy on borocarbide superconductors.

The growth of quasiparticle conduction with field is remarkably fast. At low fields,
it is linear in field (see inset of Fig.~3) and given by
\begin{equation}
\frac{\kappa_0}{T} \simeq \frac{L_0}{\rho_0} \frac{H - H_{c1}}{H_{c2}}~~~,
\end{equation}

where $\rho_0$ is the zero-field normal-state resistivity.
In some sense, this suggests that a quasiparticle in the vortex state conducts heat just
as well as in the normal state (where $\frac{\kappa_0}{T} = \frac{L_0}{\rho_0}$), 
if the density of 
quasiparticles is assumed to grow roughly as $B/H_{c2}$.

It is striking to note that quasiparticles in this $s$-wave superconductor
conduct better than they do in a typical $d$-wave superconductor.
Indeed, from calculations by Vekhter and Houghton \cite{Vekhter}, 
\begin{equation}
\frac{\kappa_0}{T} < \frac{\kappa_N}{T} \frac{H}{H_{c2}}~~~,
\end{equation}
for all $H$ in the vortex state of a $d$-wave superconductor.
Note however that the $H$ dependence is different:
it is sublinear for $d$-wave at low $H$, largely due to the 
$\sqrt{H}$ dependence of the density of states, whereas we find a linear 
dependence at the lowest fields, perhaps even going to super linear (see Fig.~3).

In trying to explain the surprisingly good conduction, it is natural to invoke
a highly anisotropic gap function, and
ascribe the excellent heat transport to the field-induced excitation of quasiparticles across
a very small gap in the basal plane. 
Izawa {\it et al.} have recently interpreted the $\sqrt{H}$ dependence of the specific heat in
YNi$_2$B$_2$C \cite{Izawa} 
in terms of a Doppler shift of the quasiparticle spectrum akin to what is seen
in $d$-wave superconductors \cite{Vekhter,Kubert} but applied in this case to a highly anisotropic
$s$-wave gap, with a small minimum gap $\Delta_{\rm min}$. 
This scenario is unsatisfactory in several respects.
First, it requires a huge gap anisotropy in a Fermi liquid that is otherwise rather isotropic. 
Indeed, because quasiparticle conduction starts right at $H_{c1}$, the average 
Doppler shift energy $E_H \simeq \Delta_0 \sqrt{H/H_{c2}}$ \cite{Kubert} at $H_{c1}$
must be greater than the minimum gap: 

\begin{equation}
\Delta_{\rm min} < E_H(H_{c1}) \simeq \Delta_0 \sqrt{H_{c1}/H_{c2}} = \Delta_0 / 10 ~~.
\end{equation}
This factor of 10 or more in gap anisotropy should be compared to the very small 
mass tensor anisotropy ({\it e.g.} from $H_{c2}$), equal to 1.01 in YNi$_2$B$_2$C 
and 1.16 in LuNi$_2$B$_2$C \cite{Kogan}.
Secondly, 
as mentioned above, it is hard to see how phonon conductivity could rise to a huge peak
at 5 K if there was significant thermal excitation of quasiparticles for $T < 5$~K,
as there would be with $\Delta_{\rm min} \simeq 1$~K.
The third problem is that we observe the same excellent quasiparticle transport  
at very low fields in a crystal of \lu\ doped with 9\% Co impurities.
The doping causes the scattering rate to be
7 times larger than in the pure crystal, so that 
$\hbar \Gamma \simeq 3.5 \Delta_0$ and $l_N \simeq \xi$. 
Such strong impurity scattering should essentially wipe out all anisotropy in the
 gap function \cite{Borkowski}, yet we observe the same transport behaviour 
in the vortex state of both pure and impure samples.

This last fact also argues against the speculation by Nohara {\it et al.} \cite{Nohara} that the field
dependence of the specific heat in YNi$_2$B$_2$C goes from $\gamma(H) \sim \sqrt{H}$ to linear in $H$
upon doping with 20\% Pt because quasiparticles go 
\begin{figure}
\resizebox{\linewidth}{!}{\includegraphics{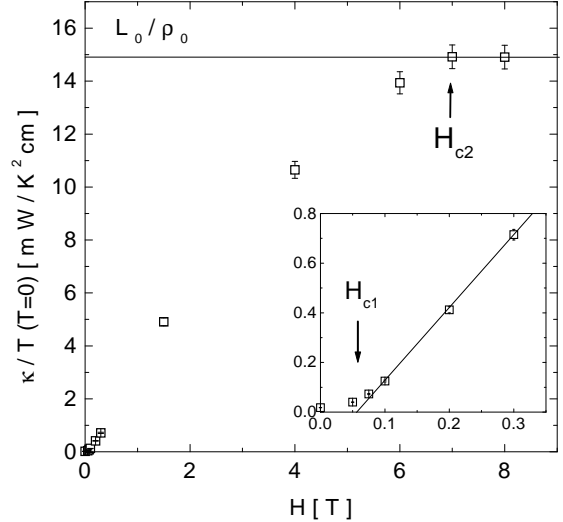}}
\caption{Magnetic field dependence of the electronic thermal conductivity
  $\kappa/T$ at $T \to 0$. The current is along the $a$-axis [100] of the
  crystal and perpendicular to the applied field. 
  The arrow gives the value of $H_{c2}$ obtained from resistivity.
  Note the rapid rise in $\kappa/T$ and the saturation above $H_{c2}$ at a value
  exactly given by the Wiedemann-Franz law.
  {\it Inset}: zoom at low fields, where a linear growth is seen to extrapolate
  precisely at the independently measured $H_{c1}$.}
\end{figure}
from being delocalized (coherent
hopping between adjacent vortices) to being localized (incoherent hopping).
We are therefore led to regard the presence of delocalized quasiparticles as a property
of the vortex lattice state and not as the result of nodes or deep minima in the gap function.
To the best of our knowledge,  
a proper calculation of the thermal conductivity in the vortex lattice state of an $s$-wave
superconductor at low fields has never been reported.
It would have to start from the quasiparticle band structure of the vortex lattice, such as 
calculated by Yasui and Kita \cite{Yasui}. For an isotropic $s$-wave gap
and a hexagonal vortex lattice at $B = 0.1 H_{c2}$,
these authors find the two lowest energy bands to be at an energy of $\Delta_1 \simeq \Delta_0
/ 2 k_F \xi$ above $E_F$ and have very low dispersion, with a band velocity no greater
than $v \simeq (a \Delta_0 / \hbar \pi) / 100$.
These two factors conspire to reduce the expected heat conduction relative to the normal state
very considerably, and it may be difficult to reconcile our results with such an energy spectrum.
First, the existence of a (mini)gap $\Delta_1$ should give rise to an 
exponential decrease of $\kappa/T$ with temperature for $k_B T < \Delta_1$. 
A rough estimate gives $\Delta_1 \simeq 300$~mK, using $k_F \simeq \pi / 2a$ and $\xi \simeq 
100$~\AA, yet we observe no decrease down to 70 mK.

Even if we assume that the lowest bands are somehow thermally populated, it is still difficult
to see how they could lead to such excellent conduction, given how flat they are.
Conductivity is weighted by the square of the carrier velocity, or band dispersion, which alone should
suppress $\kappa/T$ by a factor $(v/v_F)^2$ relative to the normal state. This implies
a reduction by a factor $2 \times 10^{7}$ at $B = 0.1 H_{c2}$.
At that field, $\kappa_e$ is only down by a factor 10.
It remains to be calculated whether the transport relaxation time $\tau$ in the vortex state
is sufficiently enhanced to compensate for that factor of $10^6$.

Theoretical expectations of a quantum phase transition from thermal metal to thermal insulator
as the vortex separation is increased by going to lower fields are not confirmed in our experiments.
The appearance of a gap at a quantum level crossing transition, proposed by
Te\u{s}anovi\'{c} and Sacramento \cite{Tesanovic}, could conceivably have been missed
if one postulates that the minigap is in fact smaller than our estimate of $\Delta_1 \simeq 300$~mK,
by a factor of 4 (or more).
The localization of quasiparticles as a result of disorder, discussed by Vishveshwara {\it et al.}
\cite{Vishveshwara}, may simply not occur in the samples we studied (both pure and Co-doped)
because the level of disorder is not high enough.

In summary, highly delocalized quasiparticles are present in \lu\
in magnetic fields as low as $H_{c1}$, in sharp
contrast with the standard picture of quasiparticle states bound to the
vortex cores at very low fields. 
We believe that this remarkably good quasiparticle conduction is an intrinsic
property of the vortex lattice state as opposed to being caused by a hugely
reduced gap in the basal plane. Calculations of heat transport in the vortex lattice state
are needed to explore the apparent compensation between the very low band dispersion
(or quasiparticle velocity) and the transport relaxation time.
Measurements on other superconductors are necessary in order to determine whether
our findings are peculiar to borocarbides or a more generic property of $s$-wave superconductors.


We are greatful to Z. Te\u{s}anovi\'{c}, S. Vishveshwara, T. Senthil, 
M.P.A. Fisher and
Michael Walker for stimulating discussions. 
This work was supported by the Canadian Institute for Advanced
Research and funded by NSERC of Canada. E. B. and C. L. acknowledge
the support of Walter C. Sumner Fellowships. L. T. acknowledges the
support of a Premier's Research Excellence Award from the Government
of Ontario. Ames laboratory is operated for the U.S. Department of
Energy by Iowa State Univesity under contract No. W-7405-ENG-82.


\end{document}